\documentclass{iaus}
\usepackage{graphicx}

\title[Triggered Star Formation] 
{Triggered Star Formation in the Environment of Young Massive Stars}

\author[Gritschneder et al]   
{Matthias Gritschneder$^1$, T. Naab$^1$, F.Heitsch$^2$, A.Burkert$^1$}

\affiliation{$^1$ Universit\"ats-Sternwarte M\"unchen, Scheinerstr.\ 1,
  D-81679 M\"unchen, Germany \\[\affilskip] $^2$ Dept. of Astronomy,
  University of Michigan, Michigan, United States}

\pubyear{2006}
\volume{237}  
\pagerange{1--4}
\date{?? and in revised form ??}
\jname{Triggered Star Formation in a Turbulent ISM}
\editors{ B. G. Elmegreen \& J. Palou\v{s}, eds.}
\begin{document}

\maketitle

\begin{abstract}
Recent observations with the Spitzer Space Telescope show clear evidence that star formation takes place in the surrounding of young massive O-type stars, which are shaping their environment due to their powerful radiation and stellar winds. In this work we investigate the effect of ionising radiation of massive stars on the ambient interstellar medium (ISM): In particular we want to examine whether the UV-radiation of O-type stars can lead to the observed pillar-like structures and can trigger star formation. 
We developed a new implementation, based on a parallel Smooth Particle
Hydrodynamics code (called IVINE), that allows an efficient treatment of the effect of ionising radiation from massive stars on their turbulent gaseous environment. Here we present first results at very high resolution.
We show that ionising radiation can trigger the collapse of an otherwise stable molecular cloud. The arising structures resemble observed structures (e.g. the pillars of creation in the Eagle Nebula (M16) or the Horsehead Nebula B33). Including the effect of gravitation we find small regions that can be identified as formation places of individual stars. We conclude that ionising radiation from massive stars alone can trigger substantial star formation in molecular clouds.
\end{abstract}

\firstsection 
\section{Overview}
In the surroundings of hot OB-Associations filamentary
substructures on different scales are observed (see
e.g. \cite{2002ApJ...565L..25S} and references therein). As
observational resolution increases, more and more sub-millimeter sources, which
could trace the birth of future stars are detected (e.g.
\cite{2006MNRAS.369.1201W}). It has long been suggested that radiation
driven implosion of molecular clouds can explain the morphology and
the star formation in these regions (e.g. \cite{1995ApJ...451..675E}). 

Recent simulations (see
e.g. \cite{2006ApJ...647..397M},
\cite{2005MNRAS.358..291D}, \cite{2003MNRAS.338..545K}) demonstrate the 
importance of massive stars for the subsequent evolution of their parental
molecular clouds. The ionising radiation is a vital ingredient to
understand the disruption of molecular clouds and their star formation
efficiency.

Our goal is to investigate the morphology of molecular clouds and the
formation of protostars in much greater detail. To do so we use very high
resolution simulations of a small region of a molecular cloud ionised
by a massive nearby star. 

\section{Numerical Method}
We use the prescription for ionising UV-radiation of a young massive
star proposed by 
\cite{2000MNRAS.315..713K}.
The ionisation degree $x$ is related to the hydrodynamical quantities
by an approximation for the resulting temperature of a partly
ionised gas
\begin{equation}
T = T_{ion} \cdot x + T_{cold} \cdot (1-x).
\end{equation}
$ T_{cold} $ is the initial temperature of the cold, unionised
gas and  $ T_{ion} $ is the temperature of the ionised gas.

To treat the hydrodynamic and gravitational evolution we use a parallel
smoothed particle hydrodynamics (SPH) code called VINE (Wetzstein
\etal, in prep.). Its Lagrangian nature renders it
extremely adept to cover several orders of magnitude in density and
time, which is 
important to follow local gas collapse. We assume plane-parallel UV-irradiation
of the simulated area, mimicking a radiation source sufficiently far
away such that its distance
is larger than the
dimensions of the area of infall. To couple ionisation to
hydrodynamics we use a flux conserving ray-shooting algorithm. A
two dimensional grid is superimposed on the area of interest. Along
each of the thereby created bins, the optical depth is calculated.
The size of each bin, i.e. the grid resolution is defined by the
volume each SPH-particle occupies. This guarantees
that the
density information given by the SPH-formalism is transformed to the
calculation of radiation correctly.

This implementation is fully parallelised. We call it IVINE (Ionisation+VINE).

\section{Numerical Tests}
A standard test for numerical implementations of ionizing radiation
has been proposed by 
\cite{1994A&A...289..559L}. It deals with the steady propagation of an
ionisation front: a box of constant density is exposed to a
time-dependent ionising source. The initial conditions were chosen to be
$n_0=100cm^{-3}$ and $T_{cold} = 100K$ to enable a direct comparison to
their results. The flux increases linearly with
time, starting at zero: $\mathrm{d}J/\mathrm{d}t =
5.07\cdot10^{-8}cm^{-2}s^{-2}$. The 
recombination parameter $\alpha_B$ is set to $\alpha_B=
2.7\cdot10^{-13}cm^3s^{-1}$. The ionised temperature is
$T_{ion}=10^4K$. At the beginning a small fraction of the box is
ionised. Due to the higher temperature of the ionised gas a shock
front evolves. This shock front moves at a constant speed through
the box. The analytical solution provides the exact position and speed of
the front as well as of the ionised gas at any given
time. 
We find a very good agreement with the analytical solution, the
results are shown in Table \ref{leflochtbl}. $n_c$ and $n_i$ denote the
number density of the compressed layer and of the
ionised gas, $v_i$ and $v_s$ are the velocities of the ionisation front
and the shock front respectively.

\begin{table}
\begin{center}
\caption{Comparison of analytical and numerical results for the Lefloch
  test.\label{leflochtbl}}
\begin{tabular}{ccccc}
\hline
& Analytical & \cite{1994A&A...289..559L} & \cite{2000MNRAS.315..713K} & IVINE  \\
\hline
$n_c(cm^{-3})$ & 159 &169 & 155 & 156.8\\
$n_i(cm^{-3})$ & 0.756 & 0.748 & 0.75 & 0.756\\
$v_i(km s^{-1})$ & 3.48 &3.36 & 3.43 & 3.41\\
$v_s(km s^{-1})$ & 3.71 & 3.51 & 3.67 & 3.63\\
\hline
\end{tabular}
\end{center}
\end{table}

\section{First application: ionisation of a turbulent ISM}
\subsection{Initial conditions}
\begin{figure}
\centering
\includegraphics[width=12cm]{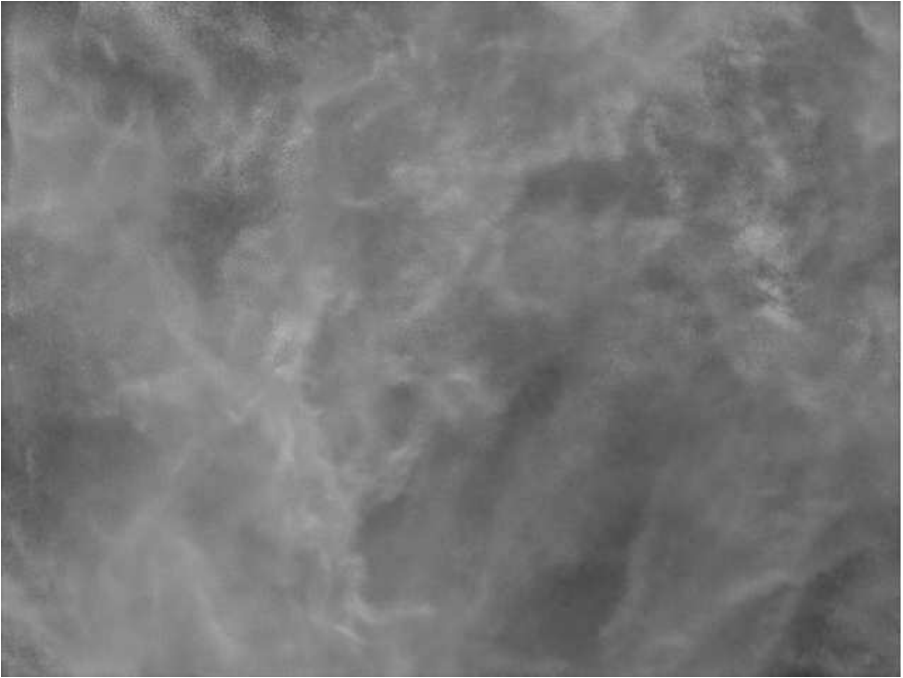}
\caption{Initial density distribution after the turbulence has decayed
to Mach 5. The box dimension are $2pc$ in each
direction.}\label{fig_init}
\end{figure}

The first high resolution simulations we performed with the new code
address the effect of ionising radiation on a box of turbulent
medium. We choose  the initial conditions to mimick observed
turbulence in the ISM. The cubic simulation domain with a volume of 
$(2pc)^3$ and a mean density of $\bar{n} =
100cm^{-3}$ is set up with a temperature of $T_{cold}=10K$. The
turbulent velocity field is set up adapting a Gaussian random field
with a steep power spectrum. The velocity field generates density
fluctuations and after a dynamical timescale a 
turbulent medium with typical velocities of Mach 5 has been generated
(see Fig. \ref{fig_init}). At this point the box is
exposed to ionising
radiation. 
The UV-radiation is impinging form the negative
x-direction with a flux $J = 8.36\cdot10^8cm^{-2}s^{-1}$. This leads to
a rapid ionisation of the first $\approx5\%$ of the cube
before the medium reacts to the increased temperature
of the ionised gas.
The simulations were performed with 2
Million particles on a SGI Altix supercomputer.
\subsection{Results}
\begin{figure}
  \centering
\includegraphics[width=12cm]{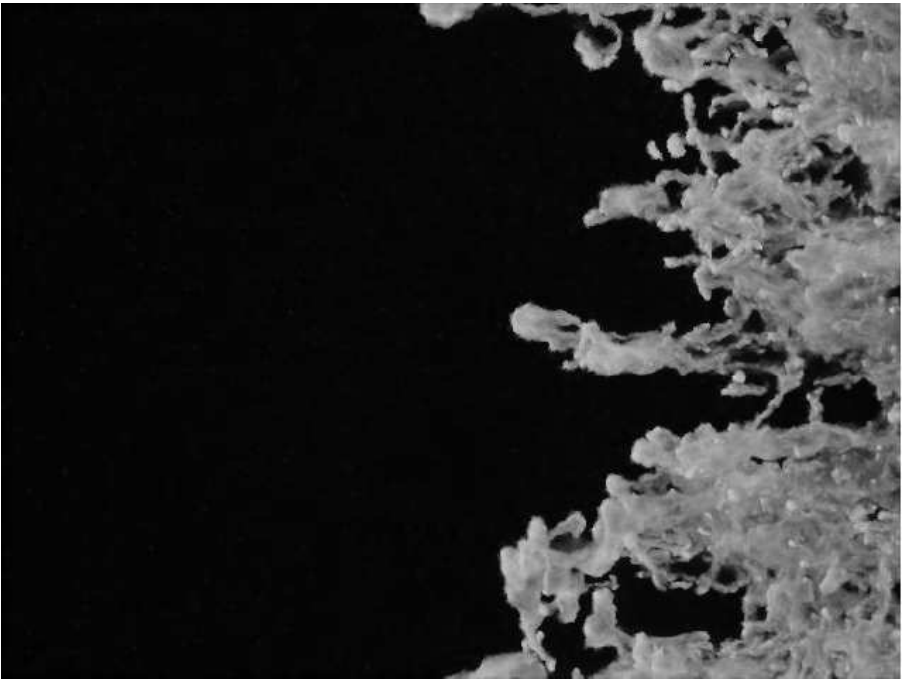}
  \caption{Final stage of evolution after $t \approx 300kyrs$: the
  morphology has clearly evolved. The size of the most prominent
  pillar is roughly $1pc$ as it is observed in e.g. M16 }\label{fig_end}
\end{figure}

The UV-radiation traces the turbulent density
distribution, reaching further into the low density regions, and less
far in the regions of high density. After a
dynamical timescale the hydrodynamics react to the increase in
temperature, shock fronts
evolve and compress the gas while at the same time increasing
the turbulent energy of
the cold gas. During this phase a typical
morphology evolves. The denser regions shadow regions behind them
whereas in lower density regimes the radiation can propagate much
further. After the first phase of maximum compression a more quiescent phase of
evaporation sets in. The densities are not as high as before, but the
structures become even more clear. This leads in the final stage to
filamentary, pillar-like 
substructures, pointing towards the source of radiation as can be seen
in Fig. \ref{fig_end}.  The structures 
contain high density regions in their tips. These a very
likely to become gravitational unstable after a free fall time.
\section{Conclusion}
We developed a fully parallel treatment for the ionising radiation of
young massive stars. Ionising Radiation alone is sufficient to explain
the morphology observed in the surroundings of hot OB-clusters. In our
simulations of turbulent ISM exposed to UV-radiation characteristic
trunks similar to the ones observed in M16 evolve. Further studies
including gravity will show whether the UV-radiation
from young massive stars is sufficient to trigger gravitational 
collapse within these pillars.

\vspace{.5cm}

\begin{acknowledgments}
M. Gritschneder is supported by the {\it Sonderforschungsbereich 375-95
Astro-Particle-Physics} of the Deutsche Forschungsgemeinschaft.
\end{acknowledgments}



\end{document}